\documentclass[aps,pre,twocolumn]{revtex4-2}
\usepackage{hyperref}
\usepackage{bm}
\usepackage{yhmath}
\usepackage{color}

\usepackage{graphicx}
\usepackage{epstopdf}

\usepackage{color}
\usepackage{mathtools}

\usepackage{amsmath,amssymb,graphicx,textcomp}
\usepackage{verbatim}

\usepackage[usenames,dvipsnames]{xcolor}
\hypersetup{colorlinks=true, linkcolor=BrickRed, urlcolor=blue!50!black, citecolor=blue!50!black}

\begin{document}

%Title of paper
\title{Anomalous transport in the soft-sphere Lorentz model}
\author{Charlotte F. Petersen}
\affiliation{Institut f\"ur Theoretische Physik, Universit\"at Innsbruck, Technikerstr. 21A, 6020 Innsbruck, Austria}
\author{Thomas Franosch}
\affiliation{Institut f\"ur Theoretische Physik, Universit\"at Innsbruck, Technikerstr. 21A, 6020 Innsbruck, Austria}

\begin{abstract}
The sensitivity of anomalous transport in crowded media to the form of the inter-particle interactions is investigated through computer simulations. We extend the highly simplified Lorentz model towards realistic natural systems by modeling the interactions between the tracer and the obstacles with a smooth potential. We find that the anomalous transport at the critical point happens to be governed by the same universal exponent as for hard exclusion interactions, although the mechanism of how narrow channels are probed is rather different. The scaling behavior of simulations close to the critical point confirm this exponent. Our result indicates that the simple Lorentz model may be applicable to describing the fundamental properties of long-range transport in real crowded environments. 
\end{abstract}

\maketitle

%%%MAIN TEXT%%%%
\section{Introduction}
Subdiffusive transport is ubiquitously  
observed in crowded heterogeneous environments,\cite{weber2010bacterial,engelke2010probing,jeon2011vivo,etoc2018non,carroll2018diffusion} which are common in natural and industrially relevant systems. Anomalous diffusion also arises in the extensively studied and highly simplified Lorentz model of crowded media, where its emergence is attributed to the percolation transition of available space, and its form is dependent on the local dynamics of diffusing particles.\cite{hofling2006localization} Simple models such as these can help in understanding the origin of anomalous transport in the more complex case of real crowded systems.\cite{hofling2013anomalous}

Situations where transport occurs in crowded environments include  catalysts,\cite{gleiter2000nanostructured,brennermacrotransport,benichou2010geometry,ben2000diffusion} ion-conductors,\cite{ben2000diffusion,voigtmann2006slow} flow in porous media,\cite{scholz2012permeability} as well as molecular sieving.\cite{gleiter2000nanostructured,han2008molecular} In the interior of crowded cells\cite{hofling2013anomalous,sokolov2012models,saxton2012wanted} the complex environment causes changes in the transport properties, which means the heterogeneous structure cannot be ignored when modeling biological reactions and macromolecular transport.\cite{hofling2013anomalous,ellis2003cell,ellis2001macromolecular,hall2003macromolecular,ellis2001macromolecular2} These changes in transport properties are usually observed through a power-law growth of the mean-square displacement, but are also manifest through persistent correlations in time, strongly suppressed time-dependent diffusion coefficients and non-Gaussian distributions of spatial displacements.\cite{hofling2013anomalous} 

In the simplest form of the Lorentz model a single point particle, referred to as a tracer, moves with Newtonian dynamics through a fixed random array of identical spherical obstacles. As the density of the obstacles increases, the space available to the tracer is reduced and the accessible space separates into unconnected regions, referred to as clusters of the void space. At a critical excluded volume, the clusters become self-similar with fractal dimension $d_{\text{f}}\approx2.53$ in three dimensions (3D).\cite{jan1998random} The cluster which spans all space is referred to as the infinite cluster.
At the percolation point of the space accessible to the tracer the model exhibits a localization transition.\cite{ben2000diffusion,jin2015dimensional,hofling2006localization} Here, the mean-square displacement increases subdiffusively for large times  like a power law\cite{ben2000diffusion}
\begin{equation}
\delta r^2(t)\sim t^{2/z}, \qquad t\to \infty,\label{eq:msd}
\end{equation}
where the dynamic universality class can be characterized by the exponent $z$. The value of this exponent has been derived through renormalization-group analysis, and depends on the percolation transition of the available space, and the distribution of transition rates through narrow channels in the obstacle matrix.\cite{straley1982non,stenull2001conductivity}

A logical extension to the original Lorentz model is to make it more applicable to naturally occurring systems. Simulations using Brownian dynamics,\cite{scala2007event,bauer2010localization,franosch2010persistent,spanner2016splitting} which are typical in biological systems of interest,\cite{einstein1905motion,von1906kinetischen} showed that the exponent $z$ depends on the dynamics of the model.\cite{spanner2016splitting} It is rather surprising that such an apparently small detail changes the universality class. 
Therefore, one must be careful when comparing anomalous transport in different models, since the mechanism of how the narrow channels are probed may be an important ingredient for the overall transport properties. So far, the consequences of using the simplified hard-sphere exclusion for the particle-obstacle interactions has not been considered.

The original model uses hard-sphere collisions between the tracer and obstacles, while softer interactions are the norm in relevant biological environments. In this work we extend the 3D Lorentz model to more realistic systems through modeling the tracer-obstacle interactions as repulsive soft interactions, instead of hard-sphere repulsions. We measure the mean-square displacement of the tracer with time around the percolation point to determine the universal dynamic exponent. This extension has been previously considered in 2D,\cite{schnyder2015rounding,yang2010anomalous,pezze2011regimes} however there the exponent describing transport is independent of the narrow channels, and as such no change from the hard-sphere case was observed.\cite{schnyder2015rounding} 
Conversely, in 3D, the value of the exponent depends on how the narrow channels are probed. In the soft model considered here, where transport occurs on a potential energy landscape, the narrow channels are high saddle points, and are conceptually different from the narrow passages of the hard-sphere model. Surprisingly, we find that the system belongs to the same university class as the hard-sphere system, albeit the arguments leading to this result no longer apply.
Our result implies that the simple Lorentz model could have applicability to real crowded media.

\begin{figure*}
\centering
  \includegraphics[height=5cm]{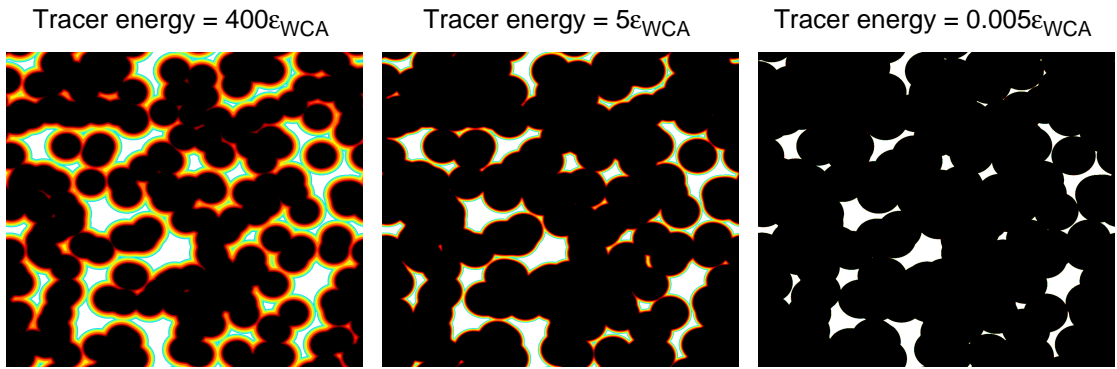}
  \caption{Schematic of the energy landscape for a two-dimensional analog.  The color scale indicates the potential at each position, where all values of the potential greater than the tracer energy are colored in black. This space is inaccessible to the tracer. As the tracer energy is reduced [left to right] the accessible space no longer percolates and all tracers are localized.}
  \label{fig:space}
\end{figure*}

\section{Simulation details}
Our model consists of a tracer moving through an array of fixed obstacles, which interact with it via a smoothed Weeks-Chandler-Anderson (WCA) potential,\cite{weeks1971role} given by
\begin{equation}
V(r)=\begin{cases}
4\epsilon_{\text{WCA}}\left[\left(\sigma/r\right)^{12}-\left(\sigma/r\right)^{6}+1/4\right]\Psi(r), & r<r_{cut},\\
0, & r\geq r_{cut},\end{cases}\end{equation}
 where $r$ is the distance of the tracer from the obstacle, and $r_{\text{cut}}=2^{1/6}\sigma$. The smoothing function is given by $\Psi(r)=\exp\left[-{h}/(r_{\text{cut}}-r)\right]$,
where we choose the width of the applied smoothing as $h=0.33 \sigma$. The length scale of the simulation is set by $\sigma$ and the energy scale by $\epsilon_{\text{WCA}}$. We keep the reduced number density of the obstacles fixed at a chosen value, $n^*=N\sigma^3/L^3=0.9$. We use $N=10^7$ randomly and independently distributed obstacles, resulting in a box size $L=223\sigma$.

The space available to the tracer is controlled through varying the energy $E$ of the tracer. The tracer can only occupy regions of space where the potential from the surrounding obstacles is less than the input tracer energy, see Fig.~\ref{fig:space}. 

The trajectory of the tracer is simulated with a leap-frog algorithm,\cite{rapaport2004art} using a time-step of $10^{-3}\sqrt{m\sigma^2/\epsilon_{\text{WCA}}}$, where $m$ is the mass of the tracer. Key simulations were repeated with a time-step one tenth the length and no change in the results was observed, demonstrating that the time-step size is sufficiently small. We carefully checked that the energy along trajectories is conserved. Times are reported in units of $t_0=\sigma(m/2E)^{1/2}$. Periodic boundary conditions are employed throughout.  

For each trajectory simulated, the initial tracer position is selected randomly, with the constraint that the potential must be lower than the energy of the tracer. The magnitude of the initial velocity is then determined from the potential at the initial position, and the direction is chosen randomly. An alternative choice would be to select the initial positions according to an equilibrium microcanonical ensemble. Simulations at the percolation point were repeated with this initialization choice, but no change in the results was observed, so we infer that the averaged long-time dynamics are not sensitive to the initial position. To improve statistics, positions along the length of the trajectory are sampled with a multiple-tau correlator algorithm\cite{berne1976dynamic,schatzel1988photon} and ten moving-time origins are used per run, separated by one percent of the total simulation time. At least 600 trajectories are simulated for each value of $E$, each with an independent obstacle configuration. For values close to the percolation point up to 1,100 trajectories are simulated. 

\section{Results and discussion}
\subsection{All-cluster dynamics}
\begin{figure}
 \centering
 \includegraphics[width=1.\linewidth]{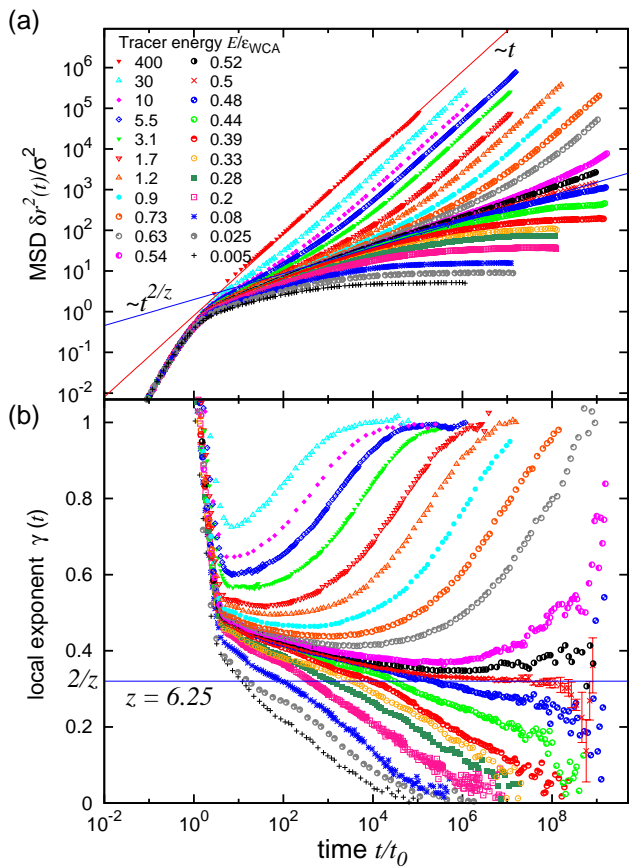}
 \caption{(a) MSD as a function of time for simulations with different tracer energy. The solid red line is proportional to time, and serves as a guide to the eye. The solid blue line indicates a power law $\sim t^{2/z}$ with the exponent $z=6.25$. (b) Local exponent of the MSD for each value of $E$, calculated numerically as the derivative of the MSD plot. Error bars on the $E_c=0.50\epsilon_{\text{WCA}}$ data are calculated by splitting the 1100 trajectories into 10 independent sets, and calculating the MSD and local exponent in each case. The standard error is plotted as the $y$-error bars. The solid blue line indicates the exponent $z=6.25$. 
}
 \label{fig:msd}
\end{figure}

We calculate the mean-square displacement (MSD) as $\delta r^2(t):=\left<|\mathbf{R}(t)-\mathbf{R}(0)|^2\right>$, where $\mathbf{R}(t)$ is the position of the tracer at time $t$ and $\left< . \right>$ indicates an average over time and different realizations of the obstacle configuration. At small values of $E$, the MSD saturates at long times, implying that all simulation realizations are localized in clusters of finite size. At large values of $E$, more space is available to the tracer, and some sampled clusters span the whole length of the simulation box. The dynamics in these clusters is diffusive. Since the initial position of the tracer is selected randomly, both spanning clusters and finite clusters are sampled by the tracers. When averaging over these diffusive and confined tracer trajectories, heterogeneous diffusion is observed, evidenced by the proportionality of the MSD with time, seen in Fig.~\ref{fig:msd}(a). 

At a certain intermediate value of $E$, the dynamics are neither localized nor diffusive. Rather, they can be described by a power law, Eq.~\eqref{eq:msd}. The MSD data can be analyzed more thoroughly, by using the local exponent,
\begin{equation}
\gamma(t):=\frac{\textnormal{d log}(\delta r^2(t))}{\textnormal{d log}(t)}.\label{eq:localexp}
\end{equation}
At the critical point the local exponent approaches a constant value at long time $\gamma(t\to \infty) = z$. By calculating the derivative numerically, we identify the critical energy as $E_c=0.50\epsilon_{\text{WCA}}$, seen in Fig.~\ref{fig:msd}(b). Due to the noise in this data, it is difficult to conclusively identify the value of the exponent $z$.

\subsection{Percolating-cluster dynamics}
The dynamics of particles confined to the infinite cluster are also anomalous, with their MSD following a power law 
\begin{equation}
\delta r_{\infty}^2(t)\sim t^{2/d_{\text{w}}}, \qquad t\to \infty,\label{eq:msd_inf}
\end{equation}
with an exponent $d_{\text{w}}$, referred to as the walk dimension. As clusters are self-similar  the walk dimension can be related to the dynamic exponent $z$ via\cite{ben2000diffusion}
\begin{equation}
z=2d_\text{w}/(2+d_\text{f}-d),\label{eq:zdw}
\end{equation}
where $d$ is the spatial dimension. Better statistics can be obtained placing tracers deliberately on the infinite cluster, therefore we focus on $\delta r^2_\infty(t)$ and the walk dimension $d_{\text{w}}$.

In the original Lorentz model, the percolating cluster can be identified from a Voronoi tessellation of the obstacle positions.\cite{schnyder2015rounding} However, for our system this approach is not possible, because the connectedness of the percolating void space depends not only on the distance of each point in space to the nearest obstacle, but rather the distance to all interacting obstacles; if the sum of the potential of these obstacles is higher than the tracer energy, the point is not accessible to the tracer. We instead extract the infinite cluster from the maximum displacement of the tracer in the  simulations.\cite{hofling2011anomalous} We initialize 10 simulations from the same initial tracer position, with different velocity directions, and calculate the trajectories for a time $10^9t_0$. All of these trajectories are clearly on the same cluster. A cluster is taken as infinite if the total maximum displacement from the 10 trajectories is larger than the box size. We found it was necessary to use 10 separate trajectories to ensure good sampling of each cluster. Following the approach used in non-equilibrium molecular dynamics,\cite{ciccotti2016non} we generate new trajectories by taking positions on the original 10 trajectories as new initial points, and reassign the velocity direction randomly. From the 117 clusters sampled, 17 infinite clusters were identified, and a further 200 trajectories were simulated for each cluster, with the starting positions taken from the original trajectories at equally spaced time intervals. 

\begin{figure}
 \centering
 \includegraphics[width=1.0\linewidth]{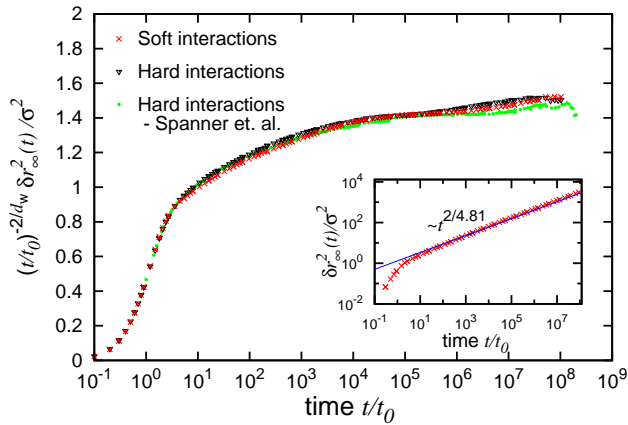}
 \caption{MSD of tracers with $E_c=0.50\epsilon_{\text{WCA}}$ on the percolating cluster, rectified with a power law given in Eq.~\eqref{eq:msd_inf}, plotted with red crosses. The MSD from hard-sphere dynamics simulations using the same method to identify the percolating cluster is plotted in black triangles. The hard-sphere results from Ref~\cite{spanner2016splitting}, where the percolating cluster is found explicitly with the Veronoi tessellation are included for comparison as green points.  Insert: MSD of the $E_c=0.50\epsilon_{\text{WCA}}$ tracers on the percolating cluster as a function of time. The solid blue line is a power law $\sim t^{2/d_{\text{w}}}$ with $d_{\text{w}}=4.81$.
 }
 \label{fig:perc}
\end{figure}

The MSD for trajectories generated in this way follows a power law, consistent with the hard-sphere exponent of $d_\text{w}=4.81$, plotted in the insert of Fig.~\ref{fig:perc}. The exponent is best confirmed by rectifying the MSD data with the expected power law, $\delta r_{\infty}^2(t)t^{-2/d_{\text{w}}}$, which approaches a constant, indicating the data is consistent with this exponent, see Fig.~\ref{fig:perc}.
Using the scaling relation in Eq.~\eqref{eq:zdw}, we see that the equivalent all-cluster average exponent $z=6.25$ is consistent with our data at the critical energy, included in Fig.~\ref{fig:msd}(b). 

We can validate this new method of identifying the percolating cluster by applying it to the hard-sphere Lorentz model, and comparing the result to  that known from the Voronoi tessellation method.\cite{spanner2016splitting} The results from our method applied to hard-sphere dynamics overlap with the soft-sphere dynamics data, see Fig.~\ref{fig:perc}. Both of these results are very close to the known hard-sphere behavior on the percolating cluster, also included in Fig.~\ref{fig:perc}, demonstrating that our method successfully identifies the percolating cluster.

\subsection{Scaling laws}
Scaling arguments\cite{ben2000diffusion,gefen1983anomalous,kertesz1984region} predict power-law singularities in the diffusion coefficient and localization length of tracers with an energy close to $E_c$.

\begin{figure}
 \centering
 \includegraphics[width=1.0\linewidth]{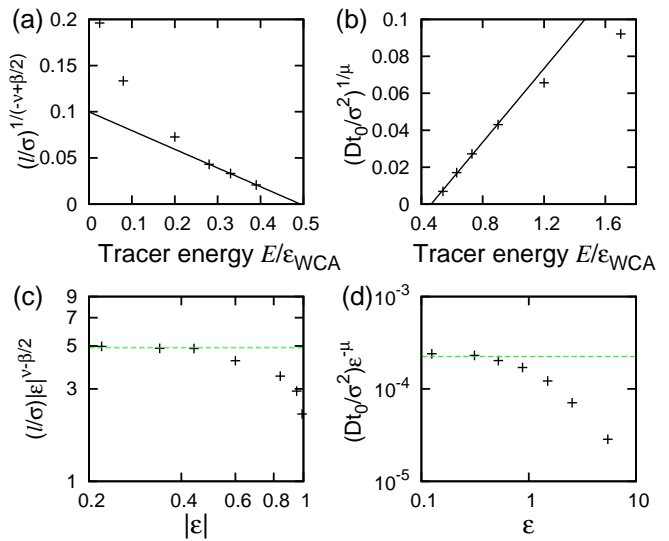}
 \caption{(a) Power-law behavior of the localization length $l$ of tracers below the critical energy. Solid line is the best fit for data close to the critical point, used to calculate the value of $E_c$ as the intercept with the $x$-axis. (b) Power-law behavior of the diffusion coefficient $D$. Solid line is the best fit close to $E_c$. (c) Rectification plot of $l$ with the expected power law. The green line is a guide for the eye of constant value. (d) Rectification plot of $D$.}
 \label{fig:lD}
\end{figure}

For its structural properties, continuum percolation belongs to the same universality class as percolation on a lattice,\cite{elam1984critical,kerstein1983equivalence} with a power-law growth of the correlation length\cite{ben2000diffusion} governed by the exponent $\nu\approx 0.88$. The maximum value of the MSD of tracers below the critical energy is purely a function of cluster size. The scaling of this size close to the transition is dependent on the universal exponents describing the geometry of the percolation network. As the percolation point is approached, the localization length is given by the mean cluster radius (radius of gyration),\cite{hofling2006localization} which scales as $l\sim |\epsilon|^{-\nu+\beta/2}$ where $\epsilon=(E-E_c)/E_c$ is a dimensionless separation parameter from the critical tracer energy. The geometric exponent\cite{ben2000diffusion} is then given by $\nu-\beta/2=0.68$. For simulations below the critical tracer energy the localization length can be read off Fig.~\ref{fig:msd}(a) as $\delta r^2(t\to \infty)=l^2$ at long times. Plotting $l^{1/(-\nu+\beta/2)}$ vs. $E$ will yield a straight line if this power law is an accurate description. In Fig.~\ref{fig:lD}(a), we confirm that the relationship holds for values close to the critical energy. From this plot we can also calculate the critical tracer energy from the $x$-intercept, finding a value $E_c=0.50\epsilon_{\text{WCA}}$, consistent with our earlier observed value. 

Above the critical energy, the time-dependent diffusion coefficient
\begin{equation}
D(t):=\frac{1}{6}\frac{d}{dt}\delta r^2(t) ,
\end{equation}
approaches a constant at long times,  yielding the diffusion constant $D:= D(t\to\infty)$. Then close to percolation the suppression of transport
is described by a power law\cite{ben2000diffusion} $D\sim \epsilon^\mu$, with a universal exponent $\mu=(z-2)(\nu-\beta/2)=2.88$ dependent on the dynamic exponent $z$. Our data is consistent with the hard-sphere exponent, shown in Fig.~\ref{fig:lD}(b). Here we identify a critical tracer energy of $E_c=0.48\epsilon_{\text{WCA}}$, very close to the values found earlier. A more sensitive method to verify the exponent is to plot $D\epsilon^{-\mu}$ against $\epsilon$, where $\epsilon$ is calculated with $E_c=0.48\epsilon_{\text{WCA}}$, shown in Fig.~\ref{fig:lD}(d). We see that the power law holds, demonstrating consistency of the diffusive simulations with the value of $z$ found at the critical energy. The same approach is used to sensitively confirm the geometric exponent in Fig.~\ref{fig:lD}(c).

\begin{figure}
 \centering
 \includegraphics[width=1.0\linewidth]{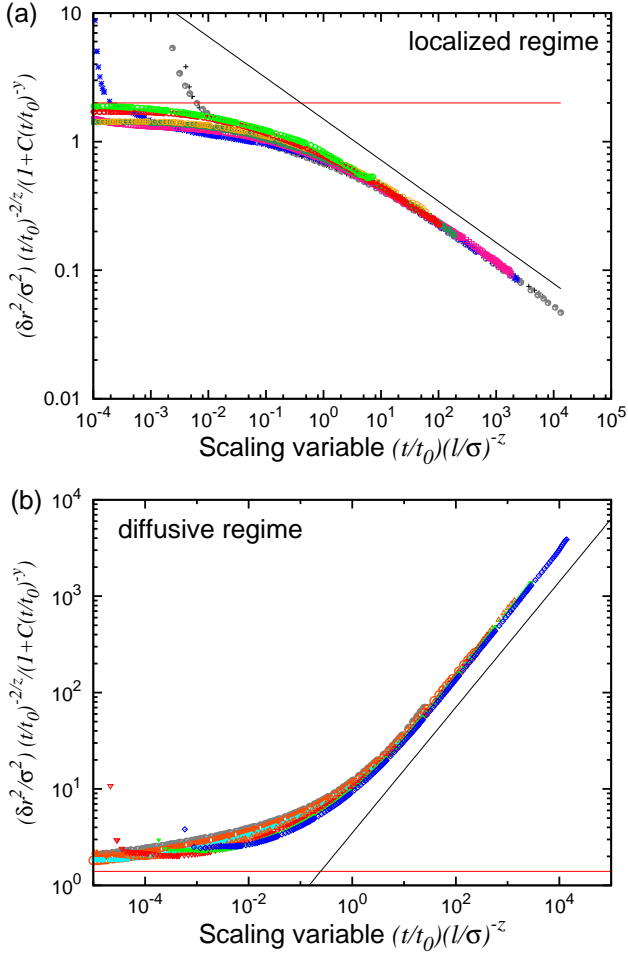}
 \caption{Scaling form of the MSD, with corrections to scaling included, plotted as a function of the scaling variable $\hat{t}$. (a)  Tracers with energy below the critical value. The red line serves as a guide to the expected constant critical asymptote at low $\hat{t}$, corresponding to $\delta\hat{r}_{\pm}^2(\hat{t})\simeq$ const. The black line corresponds to the large $\hat{t}$ asymptote, $\delta\hat{r}_+^2(\hat{t})\sim \hat{t}^{-2/z}$. (b)  Tracers with energy above the critical value. The black line corresponds to the large $\hat{t}$ asymptote, $\delta\hat{r}_-^2(\hat{t})\sim \hat{t}^{1-2/z}.$ Data colors and symbols as in Fig.~\ref{fig:msd}.}
 \label{fig:scalcorr}
\end{figure}

As in the original Lorentz model\cite{hofling2006localization} we can analyze the dynamic scaling properties by considering the scaling form of the MSD. From the scaling of the van Hove correlation function close to the critical point, the scaling form of the MSD is derived as\cite{kertesz1983properties}
\begin{equation}
\delta r^2(t; \epsilon)= t^{2/z}\delta \hat{r}_{\pm}^2(\hat{t}),\label{eq:scale_msd}
\end{equation}
where $\hat{t}\sim tl^{-z}$. For data in the diffusive regime, we calculate the scaling variable using $l=|\epsilon|^{-\nu+\beta/2}$. We account for deviations to scaling through leading-order universal corrections,\cite{hofling2006localization} given by 
\begin{equation}
\delta r^2(t; \epsilon)= t^{2/z}\delta \hat{r}_{\pm}^2(\hat{t})[1+Ct^{-y}],\label{eq:scale_msd_corr}
\end{equation}
where $y$ is the leading non-analytic correction exponent at criticality and $C$ is a constant. The exponent $y$ is calculated from the walk dimension and the correction exponent for the cluster size distribution\cite{kammerer2008cluster} $\Omega=0.64$, as $y=\Omega d_{\text{f}}/d_{\text{w}}=0.34$.   All data collapses nicely when  $t^{-2/z}\delta r^2(t)/[1+Ct^{-y}]$ is plotted against $\hat{t}$ in Fig.~\ref{fig:scalcorr}, with $C=-0.8$. This is the same amplitude $C$ used for hard-sphere dynamics previously.\cite{hofling2006localization} 

\subsection{Non-Gaussian parameter}
\begin{figure}
 \centering
 \includegraphics[width=1.0\linewidth]{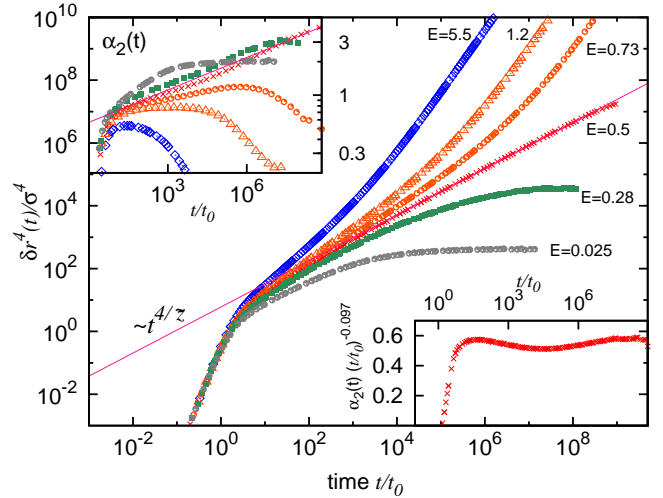}
 \caption{MQD $\delta r^4(t)$ for selected simulations with tracer energy above, below, and at the critical value $E_c=0.50\epsilon_{\text{WCA}}$. Each data set is labelled by its tracer energy, in units of $\epsilon_{\text{WCA}}$. The colors and symbols are as in Fig.~\ref{fig:msd}. The pink line indicates a power law $\sim t^{4/\tilde{z}}$ with the exponent $\tilde{z}=5.45$. Top insert: Non-Gaussian parameter for the same simulations, calculated from Eq.~\eqref{eq:NGP}. The pink line indicates a power law $\sim t^{0.097}$. Bottom insert: Non-Gaussian parameter of tracers with $E_c=0.50\epsilon_{\text{WCA}}$, rectified with the power law $t^{0.097}$.}
 \label{fig:NGP}
\end{figure}

In normal diffusion, the distribution of displacements of particles are independent, and by the central limit theorem follow a Gaussian distribution. From this, it follows that the mean-square displacement grows proportionally to $t$. This second moment of the displacement gives the value of the diffusion coefficient, while higher moments of the displacement encode no new information. In contrast, in subdiffusive motion the central limit theorem is violated, and the higher moments of the displacement provide further insight into the dynamic properties. In particular, the mean-quartic displacement (MQD), $\delta r^4(t)=\left<|\mathbf{R}(t)-\mathbf{R}(0)|^4\right>$, at the critical energy is expected to scale like a power law,\cite{hofling2006localization,spanner2011anomalous} $\delta r^4(t) \sim t^{4/\tilde{z}}$ with exponent $\tilde{z}=(2\nu-\beta+\mu)/(\nu-\beta/4)=5.45$. This is confirmed in Fig.~\ref{fig:NGP}, where we plot the MQD for selected values of the tracer energy. In the all cluster average, the distribution of displacements depends on the correlation length, as well as the distribution of cluster sizes. These length scales affect the MSD and MQD differently, resulting in different power-law exponents. 

To more sensitively measure the deviations from a Gaussian distribution we  calculate the non-Gaussian parameter,
\begin{equation}
\alpha_2(t):=\frac{3}{5} \frac{\delta r^4(t)}{[\delta r^2(t)]^2}-1. \label{eq:NGP}
\end{equation}
The prefactors are chosen in this way so that for three dimensional Gaussian transport $\alpha_2(t)=0$. For simulations with a tracer energy above or below the critical one,  the non-Gaussian parameter approaches a constant finite value at long times, see the top inset of Fig.~\ref{fig:NGP}. The limit increases as the critical energy is approached. Even in the heterogeneous diffusive regime, the non-Gaussian parameter does not decay to zero, due to the presence of  localized particles. Given that both the MSD and MQD follow power laws at the critical energy, the non-Gaussian parameter is also expected to follow a power law, $\alpha_2(t)\sim t^{4/\tilde{z}-4/z}\sim t^{0.097}$, also seen in the top insert of Fig.~\ref{fig:NGP}. As noted in the original simulation work on the hard-sphere Lorentz model,\cite{hofling2006localization} direct observation of this small exponent is a difficult task. As such, we rectify the non-Gaussian parameter data at the critical energy with the expected exponent to carefully investigate the power-law behavior. After a  short  transient, the rectified data approaches a constant, confirming the value of the exponent, seen in the bottom insert of Fig.~\ref{fig:NGP}. Notably, the non-Gaussian parameter follows a power law for over 7 decades in time, much longer than the power-law behavior of the MSD or MQD individually.  

\begin{figure}
 \centering
 \includegraphics[width=1.0\linewidth]{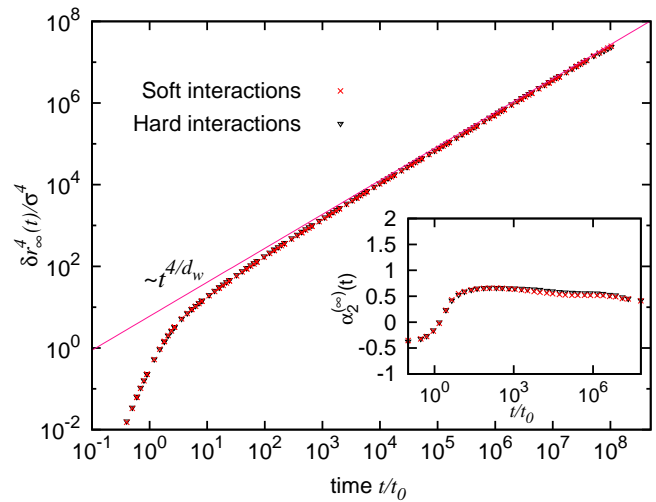}
 \caption{MQD $\delta r^4_{\infty}(t)$ of tracers located on the infinite cluster only, at the critical point. Red crosses are soft-sphere dynamics. The pink line indicates a power law $\sim t^{4/d_w}$ with the exponent $d_w=4.81$. The results for hard-sphere dynamics are included for comparison as black triangles. Insert: Non-Gaussian parameter for the same simulations.}
 \label{fig:NGPperc}
\end{figure}

We  also calculate the MQD for tracers on the infinite cluster at the critical energy.\cite{spanner2011anomalous} It follows a power law $\delta r^4_{\infty}\sim t^{4/d_w}$, seen in Fig.~\ref{fig:NGPperc}. Since by construction these tracers are on the infinite cluster, the only length scale determining the distribution of displacements is the correlation length. As such, the MQD power-law exponent is the same as the MSD exponent, $d_w$. We have included a direct comparison to our hard-sphere simulations on the infinite cluster, and find that the results overlap. For these simulations we can also calculate the non-Gaussian parameter. Since the MSD and MQD are described by the same exponent, the non-Gaussian parameter does not diverge, but attains at a finite value, see the insert of Fig.~\ref{fig:NGPperc}, in contrast to the all-cluster simulations at the critical point.

\subsection{Interpretation of the exponent}
It is instructive to reconsider the arguments from the Lorentz model for  the walk dimension $d_{\text{w}}$ as derived within renormalization group. The analysis shows that critical transport can be dominated by the transport through narrow channels.\cite{straley1982non} In 3D these channels are formed when three obstacles are positioned close together, with only a small gap between them for the tracer to pass through. If the transition rate $\Gamma$ through narrow channels  is power-law distributed  $\rho(\Gamma)\sim \Gamma^{-\alpha}$ for $\Gamma \to 0$, then the exponent $\alpha$ describing the transition rates  determines the walk dimension by the hyperscaling relation\cite{straley1982non,stenull2001conductivity}
\begin{equation}
d_{\text{w}}=\textnormal{max}\{d_{\text{w}}^{\text{lat}},d_{\text{f}}+[\nu(1-\alpha)]^{-1}\}.\label{eq:dw}
\end{equation}
If after repeated coarse graining, all open channels renormalize to the same transition rate, the situation reduces to lattice percolation, where a fixed rate is assumed from the outset, and the exponent is given by   $d_{\text{w}}^{\text{lat}} = 3.88$ (3D),  the universal exponent for Boolean random-resistor networks.\cite{ben2000diffusion} In contrast, for sufficiently high $\alpha$, the tails persist even after coarse graining and the narrow channels determine the universality class.

In the original Lorentz model an exponent, $\alpha=1-(d-1)^{-1}=1/2$ in 3D, was predicted by Machta and Moore\cite{machta1985diffusion} and later corroborated through computer simulations.\cite{hofling2006localization,hofling2007crossover,franosch2011space,spanner2011anomalous,spanner2013dynamic,spanner2016splitting} This expression can be derived from the geometry of the narrow channels, by assuming the transition rate through a channel is proportional to its cross-sectional area.\cite{spanner2016splitting} When using Brownian dynamics, the exponent is given by $\alpha=(d-5/2)/(d-3/2)$, which is derived by also taking into consideration the effective length of the narrow channel.\cite{halperin1985differences,spanner2016splitting} Our measured dynamic exponent $d_w$ is the same as for hard-sphere Newtonian dynamics, implying that the exponent $\alpha$ is also the same, and the models are part of the same universality class.

Yet, the derivation of $\alpha$ must be different for soft spheres, since it is the energy distribution of  saddles rather than the width distribution of channels that becomes relevant. 
In the original Lorentz model, the obstacles form a maze which the tracer explores, where the narrow channels are narrow passages. The situation is different for soft interactions; the obstacles form a potential landscape, consisting of valleys and mountains. Here, the narrow channels are high mountain passes, over which the tracer only just has enough energy to cross. The natural quantity by which to characterize a narrow channel is the difference in energy between the saddle point and the tracer energy, which we  denote $\Delta E$. 
 The probability distribution $P(\Delta E)$ is anticipated to be regular for small $\Delta E$, $P(\Delta E\to 0) = \text{const.}$, similar to the width distribution in the hard-sphere model. Generically, the saddles are locally parabolas and their linear size then scales as $\sim\sqrt{\Delta E}$. The probability distribution for the transition rates $\rho(\Gamma)$ is obtained from the energy distribution by $\varrho(\Gamma)d\Gamma=P(\Delta E)d\Delta E$, once the relation between transition rates $\Gamma = \Gamma(\Delta E)$ and energy $\Delta E$ is known. Then one convinces oneself that the $\alpha=1/2$ is recovered provided $\Gamma \sim (\Delta E)^2$ for $\Delta E \to 0$. Let us speculate, how this originates from the picture of high-lying mountain passes. 
The transition rate to cross a given channel will be proportional to the phase space volume at the neck of the channel, rather than simply its cross sectional area. We can consider  position and momentum space separately. The cross-sectional area of the channel neck will scale as its linear dimension squared $\sim\Delta E$. Assuming further that the potential in all accessible parts of the channel neck contribute, the phase-space volume corresponding to every point in position space scales as $\sqrt{\Delta E}$. Therefore, one anticipates that the statistics of the narrow channels changes, with possible ramifications for the distribution of transition rates. Next we reconsider how the narrow channels are probed by the traces. 
Unlike the original hard-sphere Lorentz model, where the tracer moves with constant speed, the speed of the tracer in our system changes as it moves over the potential landscape. This means the transition rate will also be proportional to the speed of the tracer at the neck of the channel, which scales as $\sqrt{\Delta E}$. Combining these three factors suggests for  the transition rate $\Gamma\sim (\Delta E)^2$.

\section{Summary and Conclusions}
We have extended the Lorentz model towards realistic systems by replacing the hard-sphere interactions with soft interaction potentials. Through simulations at the critical point on the percolating cluster, we observe that the dynamics are anomalous, following a power law with the same exponent as the hard-sphere Lorentz model. We confirm this exponent through scaling relations with simulations at energies around the critical one. Our results demonstrate that the simple Lorentz model is relevant also for  experimental systems,\cite{sung2006lateral,sung2008lateral,skinner2013localization,schnyder2017dynamic} where soft interactions are the norm.

Here we used only independently distributed obstacles, since adding structural correlations to the original Lorentz model does not affect the universality class.\cite{spanner2016splitting,saxton2010two,cho2012effect,sung2008effect} However, recent work indicates that the local exponent describing transport may be dependent on the shape of the obstacles,\cite{priour2018percolation} thereby leading to slowly fading transient dynamics.   
As such, considering diffusion on more complex potential landscapes\cite{pezze2011regimes,robert2010anisotropic,evers2013colloids,bewerunge2016experimental,shvedov2010laser,camboni2012normal,thiel2013disentangling} could yield valuable insight particularly relevant to real crowded media, where the obstacles are unlikely to be spherical. Interestingly, our result is in contrast to what is seen in a periodic Lorentz gas, where softening the interactions drastically changes the nature of the transport.\cite{klages2019normal} 

As a further extension to our work, we expect that the combination of using soft interaction potentials with Brownian dynamics will lead to interesting new results, with particular relevance to biological systems,\cite{ellis2001macromolecular,ellis2003cell,hall2003macromolecular,kim2009effect,kim2010crowding,kwon2014dynamics,saxton2012wanted,sokolov2012models,hofling2013anomalous,zeitz2017active,feig2017crowding,oh2018lateral,stylianidou2018strong,weiss2018tale} where soft interactions and Brownian motion are typical. As in the case of adding many interacting tracers to the Lorentz model,\cite{schnyder2018crowding,horbach2010localization,horbach2017anomalous,kurzidim2009single,kurzidim2010impact,kurzidim2011dynamic,kim2010molecular,kim2011slow} the localization transition will become rounded. A similar smoothing effect might be observed for Newtonian dynamics if thermal vibrations or rearrangements\cite{sentjabrskaja2016anomalous} of the host matrix is considered. In cases where the localization transition becomes rounded, the measured exponents describing transport will vary from the Lorentz model.\cite{schnyder2018crowding} However, knowledge of the exponent from the underlying percolation transition, which we have identified here, will allow for identification of this rounding. Additionally, as the conditions studied here are approached, we expect the Lorentz model to become quantitatively applicable. This situation corresponds to a low density of diffusing particles in a rigid percolating structure, which could include the case of gas diffusion through porous rock.

In the original Lorentz model, whether narrow channels dominate or not depends on the dimension of the embedding space. It turns out that in two dimensions (2D) the universality class of random-resistor networks is recovered,\cite{bauer2010localization} while in 3D the transport is dominated by the narrow channels. This is true also for soft interaction potentials. When the constraints of the hard-sphere idealization are relaxed in 2D systems,\cite{schnyder2015rounding,yang2010anomalous,pezze2011regimes} the random-resistor universality class is recovered,\cite{schnyder2015rounding} matching the hard-sphere-interaction case. Thus, in both two and three dimensions the hard-sphere Lorentz model is in the same universality class as the corresponding soft-sphere model, but for different reasons: in 2D it is because the lattice universality class is recovered, while in 3D it is because the value of $\alpha$ happens to coincide for the flat maze and the high mountain passes.

\section*{Acknowledgements}
We thank Felix H\"ofling and Suvendu Mandal for insightful discussions. This work has been supported by the Austrian Science Fund
(FWF): I 2887. The computational results presented have been achieved in part using the HPC infrastructure LEO of the University of Innsbruck. 

%%%END OF MAIN TEXT%%%

%%%REFERENCES%%%
\bibliography{lorentz2} 

\end{document}